\documentstyle[12pt]{article}
\def\ba{\begin{eqnarray}}
\def\ea{\end{eqnarray}}

\def\lb{\label}
\def\be{\begin{equation}}
\def\ee{\end{equation}}

\begin{document}

\title{New Spin(7) holonomy metrics admitting $G_2$ holonomy reductions and M-theory/IIA dualities}
\author{S. Salur $^{1}$ and  O. Santillan $^{2}$ \thanks{%
firenzecita@hotmail.com, salur@math.rochester.edu and osantil@maths.tcd.ie} \\
{\small 1- Department of Mathematics University of Rochester,
Rochester, NY, 14627, USA.}\\
{\small 2- Hamilton Mathematics Institute and School of Mathematics Trinity College Dublin 2 Ireland}\\}
\date{}
\maketitle

\begin{abstract}
As is well known, when D6 branes wrap a special lagrangian cycle on a non compact CY 3-fold in such a way that the internal string frame metric is Kahler there exists a dual description, which is given in terms of a purely geometrical eleven dimensional background with an internal metric of $G_2$ holonomy. It is also known that when D6 branes wrap a coassociative cycle of a non compact $G_2$ manifold in presence of a self-dual two form strength the internal part of the string frame metric is conformal to the $G_2$ metric and there exists a dual description, which is expressed in terms of a purely geometrical eleven dimensional background with an internal non compact metric of Spin(7) holonomy. In the present work it is shown that any $G_2$ metric participating in the first of these dualities necessarily participates in one of the second type. Additionally, several explicit Spin(7) holonomy metrics admitting a $G_2$ holonomy reduction along one isometry are constructed. These metrics can be described as $R$-fibrations over a 6-dimensional Kahler metric, thus realizing the pattern Spin(7) $\to G_2\to$ (Kahler) mentioned above. Several of these examples are further described as fibrations over the Eguchi-Hanson gravitational instanton and, to the best of our knowledge, have not been previously considered in the literature.
\end{abstract}

\section{Introduction}
   
       Spaces of $G_2$ and Spin(7) holonomy were the only two cases of the Berger list of the possible holonomy groups for Riemanian geometries \cite{Berger} whose existence was not clear. This situation completely changed with the construction of explicit non compact examples in \cite{Bryant}-\cite{Gibbons} and the proof of the existence of compact ones given in \cite{Joyce}-\cite{Kowaleski}. Since the appearance of these works, further special holonomy metrics were found in \cite{novo1}-\cite{Gaston}. The reduction of the holonomy from $SO(7)$ or $SO(8)$ to $G_2$ or Spin(7) implies  that these metrics are Ricci flat, that is, $R_{ij}=0$ being $R_{ij}$ the Ricci tensor constructed with the special holonomy metric under consideration.  Another of their salient features is the presence of at least one covariantly constant Killing spinor $\eta$, that is, an globally defined spinor satisfying $D\eta=0$ being $D$ the standard covariant derivative in the representation of the field. If the holomy is exactly $G_2$ or Spin(7) there is only one of such spinors, in other cases the holonomy will be reduced to a smaller subgroup. In fact, the presence of a parallel spinor $\eta$ makes these spaces relevant for constructing supersymmetric solutions of supergravity theories or even vacuum solutions of superstring theories. This comes
 from the general fact  that the number of supersymmetries preserved by these solutions is related to the number of independent parallel spinors that the internal manifold admits. Compactifications of M-theory (or its low energy limit, eleven dimensional supergravity) on $G_2$ or Spin(7) spaces give  N=1 supersymmetric theories in four and three dimensions respectively. Additionally, compactifications of heterotic string theory on these spaces also provide N=1 supersymmetry in 3 and 2 dimensions respectively \cite{Pap}.  
   
        From a phenomenological point of view, compactifications of eleven dimensional supergravity over $G_2$ holonomy spaces constitute an attractive possibility, as the resulting low energy theory is four dimensional. But if the internal space is smooth then the four dimensional theory will be N=1 supergravity coupled to abelian vector fields, and no chiral matter on non abelian vector fields will appear. Nevertheless, non perturbative effects arising by singularities may generate chiral matter and non abelian gauge fields in four dimensions \cite{Becker0}. For this reason special attention was paid to $G_2$ holonomy spaces developing conical singularities.  Another motivation for studying special holonomy manifolds appears in the context of dualities, as the present understanding of the dynamics of N=1 supersymmetric theories relies partially in the existence of dual realizations of a given theory.  An old example was considered in \cite{Atiyah}-\cite{Acharya}, \cite{Vafa}  where it was shown that  type IIA propagating on the deformed conifold with D-branes and IIA on the resolved conifold with RR fluxes are dual to each other. This duality has been derived by lifting both backgrounds to purely geometrical M-theory ones with two $G_2$ holonomy manifolds  admitting an smooth interpolation.  As D-branes contains gauge fields, this allows to study infrared dynamics by means of M-theory on $G_2$ manifolds.  Other contexts in which these spaces appear are in \cite{Becker}-\cite{Becker18}.
            
          New dual descriptions involving special holonomy manifolds were reported in \cite{Minasian}. In this reference the geometries corresponding to D6 branes wrapping a supersymmetric 3-cycle in non compact Calabi-Yau and to D6 branes wrapping a coassociative 4-cycle in a non compact $G_2$ holonomy space were described from an eleven dimensional perspective. Both cases give N=1 supersymmetry. For the CY case it was shown that  if the ten-dimensional string frame is a wrapped product with an internal Kahler metric then the supersymmetry generator becomes a covariantly constant gauge spinor. This requirement, which is stronger than N=1 supersymmetry, is know as "strong supersymmetry condition" and is translated into a system of equations involving the dilaton and the RR two form. These equations are a sort of  "monopole equations" describing the special lagrangian cycles the D6 branes wrap and the back-reaction of the branes on the physical metric. The lift of these IIA solutions to eleven dimensions results in a purely geometrical background with a internal $G_2$ holonomy metric.  Similarly, for D6 branes wrapping a coassociative cycle on a $G_2$ manifold with self-dual two form field strength the ten-dimensional string frame is a wrapped product with an internal metric conformal to the $G_2$ holonomy one.  As for the previous case, the lift of these IIA solutions to eleven dimensions result in a purely geometrical background, but the internal metric is now of  Spin(7) holonomy.

        There are two inherent mathematical problems that arise in the context of the dualities mentioned above.  One is the classification of the $G_2$ holonomy metrics possessing an isometry action preserving the closed $G_2$ structure in such a way that the quotient of the 7-dimensional structure by this action is Kahler; the other is  the classification of the Spin(7) holonomy metrics with an structure preserving isometry such quotient of the 8-metric by this action gives a 7-metric conformal to a $G_2$ holonomy one. The former situation has been studied in more detail. In fact  in \cite{Ego}, one of the authors has identified these $G_2$ metrics with the ones discovered independently in \cite{Apostolov}, and the relation between the generalized monopole equation and those classifying the $G_2$ geometry in \cite{Apostolov} was pointed out explicitly. Besides, an infinite class of explicit examples were presented in \cite{Gaston}. All these examples are described as fibrations over hyperkahler metrics of the Gibbons-Hawking type. Instead, the situation corresponding to Spin(7) manifolds is less understood and one of the purposes of the present work is to study it further. A natural question is wether or not the $G_2$ metrics participating in both dualities are related. In the present work it is shown that \emph{any} $G_2$ metric admitting a 6-dimensional Kahler reduction along an isometry can be obtained by a quotient of a closed Spin(7) structure by an structure preserving isometry. We are not able to prove or reject the inverse statement. The conclusion is that the set of $G_2$ metrics obtained by reduction of a Spin(7) holonomy manifold in the way described above is \emph{equal or bigger} than the ones that admit 6-dimensional Kahler reductions.

       It is tempting to connect the mathematical and physical aspects of the dualities described above by saying that whenever a $G_2$ metric provides a dual description for D6 branes wrapping a special lagrangian cycle on a CY satisfying the strong supersymmetry conditions it also describes a configuration of D6 branes wrapping a coassociative cycle inside the $G_2$ manifold in presence of a self-dual RR two form. The fact that the corresponding backgrounds are completely determined in terms of the $G_2$ metric may suggest that those D6 brane configurations are dual to each other, and the link between them is provided by the $G_2$ structure.  This is a very interesting statement but we are still cautious for the following reason. A configuration of D6 branes wrapping a special lagrangian cycle in a CY manifold will appear only if magnetic sources for the RR two form F are present, that is, dF = N $\delta$. Consider the $G_2$ metric dual to one of such configurations. By use of the result of the present work, one can lift it to an Spin(7) metric and construct an eleven dimensional background which is the direct sum of this metric with the Minkowski one in three dimensions. Clearly, the usual Kaluza-Klein reduction along an isometry gives a IIA background with a non trivial dilaton and a self-dual RR two form F$^{'}$, and the internal metric is conformal to the $G_2$ holonomy metric. If there are delta type of sources for  F$^{'}$, then this configuration will correspond to D6 branes wrapping a coassociative cycle and the duality we are talking about seems to be plausible. But we did not find a formal argument which insures that such singularities will appear, even if they were present  for the initial configurations. One can argue that dF$^{'}$=0 everywhere, but F$^{'}$ is non trivial due to bad asymptotics at infinite. In our opinion this is not the case, up to possible pathological counterexamples. In any case, we suspect that our result  encode a very interesting class dualities between D6 brane configurations.
                
         In addition, we are able to find new $G_2$ holonomy examples not considered in \cite{Apostolov} and their lift to Spin(7) metrics. All these examples arise as fibrations over the Eguchi-Hanson
 gravitational instanton and the fiber quantities are defined over a complex submanifold of the Eguchi-Hanson space. In this situation all the fiber quantities are defined by the solution of a laplace type equation on the curved instanton metric, otherwise the underlying problem becomes non linear and in consequence, harder to solve. 
          
      The present work is organized as follows. In section 2  it is presented a system of equations describing the lift of a $G_2$ holonomy metric to an Spin(7) one, which is essentially the one considered in \cite{Minasian}. In addition, a brief characterization of the $G_2$ holonomy metrics which admit a Kahler reduction is given. In particular, it is shown that \emph{any} of these $G_2$ metrics can be lifted to an Spin(7) one by means of these equations, which is one of the main results of the present work. In section 3 some known examples of these $G_2$ metrics \cite{Apostolov}-\cite{Gaston} are presented and the lifting to Spin(7) metrics is performed explicitly. In section 4 we review a method for constructing $G_2$ holonomy metrics admitting Kahler reductions in terms of an initial hyperkahler 4-dimensional metric together with certain quantities defined over a complex submanifold of the hyperkahler manifold \cite{Apostolov}. We show that this method linearize the otherwise non linear system describing this geometry, and converts it into a Laplace type equation on the curved hyperkahler metric. We find a non trivial solution when the hyperkahler manifold is the Eguchi-Hanson gravitational instanton and construct the corresponding special holonomy metrics. In the last section we make a brief discussion of the presented results.

 \section{Spin(7) metrics admitting $G_2$ reductions}
  
\subsection{The defining equations}

    As it was mentioned in the introduction, a configuration of D6 branes wrapping a coassociative submanifold of a $G_2$ manifold in such a way
 that the field strength $F_{ab}$ satisfying the self-duality condition
\be\lb{selfdis}
4\;F_{ab}+ c_{abcd}\; F_{cd}=0,
\ee
is described in terms of a type IIA background with an internal 7-metric which, in the string frame, is conformal to a $G_2$ holonomy metric \cite{Minasian}. Here $c_{abcd}$ are the duals of the octonion multiplication constants. Any of these IIA backgrounds can be lifted to a purely geometrical solution of eleven-dimensional supergravity of the form 
\be\lb{surga}
g_{11}=g_{(1,2)}+g_8,
\ee
being $g_8$ a Spin(7) holonomy metric possessing a Killing vector preserving also the Spin(7) calibration 4-form.  Here $g_{1,2}$ is the Minkowski metric in 
three dimensions.The purpose of the section is to clarify the relation between these $G_2$ and Spin(7) holonomy metrics. 

        Consider an 8-dimensional space $M_8$ with metric
\be\lb{spin}
g_8=e^{6 f}(dz+A)^2+e^{-2 f}g_7
\ee
such that the 1-form $A$, the 7-metric $g_7$ and the function $f$ are independent on the coordinate $z$. This condition means that $V=\partial_z$ is a local Killing vector, which induce a local decomposition $M_8=M_7\times R_{z}$ if $z$ is non compact or $M_8=M_7\times U(1)_z$ if $z$ is an angular coordinate. In the following we will impose that  
$g_8$ is of Spin(7) holonomy and that $g_7$ is of $G_2$ holonomy and we will derive the consequences of this statement, with the further assumption that $V=\partial_z$ also preserve the Spin(7) structure. 

    By defining the one form $e^8=e^{3 f}(dz+A)$ the Spin(7) calibration 4-form corresponding to $g_8$ can be decomposed in the following form
\be\lb{chu}
\Omega_8=e^{8}\wedge \widetilde{\Phi}+\ast \widetilde{\Phi}.
\ee
Here $\widetilde{\Phi}$ and $\ast \widetilde{\Phi}$ are a pair of $G_2$ invariant 3 and 4 forms for the metric $e^{-2f}g_7$. As the function $f$ appearing in the expression (\ref{chu}) is z-independent, it follows that the whole 4-form (\ref{chu}) will be preserved by $V=\partial_z$. Furthermore $\widetilde{\Phi}=e^{-3f}\Phi$ and $\ast\widetilde{\Phi}=e^{-4f}\ast\Phi$ being $\Phi$ and $\ast \Phi$ certain $G_2$ invariant 3 and 4 forms for the metric $g_7$.
The four form (\ref{chu}) can be expressed in terms of $\Phi$ and $\ast\Phi$ as
\be\lb{chu2}
\Omega_8=(dz+A)\wedge \Phi+ e^{-4f}\ast \Phi.
\ee
As $g_7$, by assumption, has holonomy in $G_2$ it follows that $d\Phi=d\ast\Phi=0$. Then the condition for Spin(7) holonomy $d\Omega_8=0$ will be equivalent to the following system
\be\lb{mon}
F\wedge \Phi+ d(e^{-4f})\wedge \ast\Phi=0,
\ee
being $F=dA$. By construction $F$ is a closed two form. 

     In principle, if one start with a closed $G_2$ structure and solve (\ref{mon}) then the result is a Spin(7) holonomy metric. The problem is that,  in general, it is not easy to find 
a non trivial solution. In fact, if one starts with an arbitrary $G_2$ metric it can be hard task to guess an anzatz for $F$ and $f$ in such a way that the resulting system of equations
takes a manageable form.  Let us also note also that this system does not classify completely all the Spin(7) metrics admitting a $G_2$ holonomy reduction. Even if $d\Phi=d\ast\Phi\neq 0$ there could exist a rotation of the tetrad frame of $g_7$ such that $d\Phi^{'}=d\ast\Phi^{'}=0$ for certain new calibration forms. We will try not to classify all the possible solutions of (\ref{mon}), but instead we will find
some particular ones. The $G_2$ metrics from which we will start are an special
class of $G_2$ holonomy metrics which are defined by admitting Kahler reductions \cite{Minasian}, \cite{Gaston}, \cite{Apostolov} and \cite{Ego}. Fortunately, we will be able to solve (\ref{chu}) for all these metrics.

     Clearly, the Spin(7) metrics presented above can be extended to a purely geometrical background of the form (\ref{surga}). This background can be rewritten in the IIA form 
\be\lb{mert} 
g_{11}= e ^{-\phi} g_{10} + e ^{2\phi}(dz+H_3)^2, 
\ee 
being $V=\partial_{z}$ the corresponding Killing vector. The usual reduction to IIA supergravity gives 
\be\lb{diuha} 
g_{IIA}=\eta^{1/3} g_{(1,2)}+ \eta^{-1/9} g_{7}, \qquad F=\omega_3, 
\ee
being $g_7$ the $G_2$ holonomy metric and where the dilaton $\phi$ is defined through the relation $e^{2\phi}=\eta^{-2/3}$.  The 
seven dimensional internal part of the background (\ref{diuha}) is then \emph{conformal} to the $G_2$ metric, in accordance with \cite{Minasian} and our previous discussion. 

\subsection{$G_2$ holonomy metrics admitting Kahler reductions}

   The next step is to find non trivial solutions of the system (\ref{mon}), or equivalently, to construct non trivial Spin(7) metrics possessing an isometry such that the orbits
of the Killing vector induce a 7-dimensional metric conformal to a $G_2$ holonomy metric. As we will show below, the system (\ref{mon}) can be solved for the large 
class of $G_2$ metrics considered in \cite{Apostolov} and independently in \cite{Minasian}. These metrics always possess a Killing vector which preserve the whole $G_2$ structure such that the induced metric by taking the quotient with 
respect to this isometry is a six dimensional Kahler metric.  In \cite{Minasian} the local form of these metrics is described
in terms of "generalized monopole equations" while in \cite{Apostolov} the description is given in terms of a non linear system that we will describe below.  In addition, the analysis of \cite{Ego} show that both descriptions are equivalent. The reason for choosing the second formalism is that, as we will see, it considerably simplifies the lifting equation (\ref{mon}).  
              
   In general, if a $G_2$ holonomy metric possess an isometry preserving the whole $G_2$ structure,  then the orbits of the Killing vector induce an $SU(3)$ structure with generically non zero torsion classes \cite{Chiossi}. But if the associated $SU(3)$ structure is Kahler, then the initial $G_2$ has another isometry which commute with the former one \cite{Apostolov}. Therefore any of such $G_2$ metrics is toric from the very beginning.  In addition, it is possible to make a further reduction with respect to the second isometry and describe the $G_2$ and Kahler metrics
as fibrations over certain Kahler 4-dimensional metric which we will specify below. \footnote{Note that the lifting of these metrics to 8 dimensions by (\ref{mon}) will give an Spin(7) metric
with three commuting isometries, as the initial $G_2$ metric is toric.} 

         Let us describe schematically the local form of the $G_2$ holonomy metrics in question, further details can be found in the original reference \cite{Apostolov}. Their local form is
\begin{equation}
g_{7}=\frac{(d\alpha +H_{2})^{2}}{\mu ^{2}}+\mu \;\left( \;u\;d\mu ^{2}+%
\frac{(d\beta +H_{1})^{2}}{u}+g_{4}(\mu )\;\right).
\label{senio}
\end{equation}%
All the quantities defining $g_7$ are independent on the coordinates $\alpha$ and $\beta$, therefore (\ref{senio}) is toric with Killing vectors 
$V_1=\partial_{\alpha}$ and $V_2=\partial_{\beta}$. The metric $g_4(\mu)$ is Kahler and defined over a four manifold $M$ and it depends on $\mu$ as a parameter. It also admits a complex $\mu$-independent symplectic
2-form $\Omega =\omega_{2}+i\omega_{3}$, where being
``symplectic'' means that it is closed, $d\Omega =0$. On the other hand,
being ``complex'' implies that 
\begin{equation}\omega_{2}\wedge \omega_{2}=\omega_{3}\wedge \omega%
_{3},\qquad \omega_{2}\wedge \omega_{3}=0, \label{areva0}
\end{equation}%
and that the equation
\begin{equation}
\omega_{2}(J_{1}\cdot,\cdot )=\omega_{3}(\cdot,\cdot ).
\label{areva}
\end{equation}%
define a complex structure $J_{1}$. In other words, the Niejenhuis tensor of $J_{1}$ vanishes identically or equivalently
$J_{1}$ is integrable. The two form $\widetilde{\omega}_1(\mu)$  constructed by lowering the indices of $J_1$ with $g_4(\mu)$ is in general $\mu$ dependent
 and closed on $M$. It is also orthogonal to $\omega_2$ and $\omega_3$ with respect to the wedge product, that is
\be\lb{areva4}
\widetilde{\omega}(\mu)\wedge \omega_2=\widetilde{\omega}(\mu)\wedge \omega_3=0.
\ee
The function $u$ in (\ref{senio}) depends on the coordinates of $M$ and
on the parameter $\mu$, and is defined through the relation 
\begin{equation}
2\mu\; \widetilde{\omega}_{1}(\mu )\wedge \widetilde{\omega}_{1}(\mu )=u\;\Omega \wedge 
\overline{\Omega }.  \label{compota}
\end{equation}%
This function always exists because the wedge products in (\ref%
{compota}) are proportional to the volume form $V(g_4)$ of $g_4(\mu)$. In fact 
$$
\widetilde{\omega}_{1}(\mu )\wedge \widetilde{\omega}_{1}(\mu )=V(g_4)
$$
The forms $H_{1}$ and $H_{2}$ are defined on $M\times {\bf R}_{\mu }$
and $M$ respectively by the equations 
\begin{equation}
dH_{1}=(d_{M}^{c}u)\wedge d\mu +\frac{\partial \widetilde{\omega}_{1}}{\partial
\mu },\qquad dH_{2}=-\omega_{2},  \label{chon}
\end{equation}%
with $d_{M}^{c}=J_{1}d_{M}$. The last equation can always be solved locally as the forms $ \widetilde{\omega}_{1}$ and $\omega_{2}$ are closed.
The integrability condition associated to the first (\ref{chon}) is the evolution equation
\begin{equation}
\frac{\partial ^{2}\widetilde{\omega}_{1}}{\partial ^{2}\mu }=-d_{M}d_{M}^{c}u.
\label{ebol}
\end{equation}%
Now a theorem given in \cite{Apostolov} insures that  if the system of equations described above are satisfied then the metric (\ref{senio}) are of $G_2$ holonomy. 
This statement is not difficult to see. The calibration $3$-form corresponding to the metrics (\ref{senio}) is 
\[
\Phi =\widetilde{\omega}_{1}(\mu )\wedge (d\alpha +H_{2})+d\mu \wedge (d\beta
+H_{1})\wedge (d\alpha +H_{2}) 
\]%
\begin{equation}
+\mu \;\left( \;\omega_{2}\wedge (d\beta +H_{1})+u\omega%
_{3}\wedge d\mu \;\right) ,  \label{dale}
\end{equation}%
and the dual form $\ast \Phi$ corresponding to (\ref{dale}) is given by \cite{Gaston}
$$
\ast\Phi=\mu^2\widetilde{\omega}_{1}(\mu )\wedge d\mu\wedge(d\beta+H_1)+u \omega_{2}\wedge (d\alpha+H_2)\wedge d\mu
$$
\be\lb{dol}
+\omega_{3}\wedge (d\beta+H_1)\wedge (d\alpha+H_2)+\mu^2\widetilde{\omega}_{1}(\mu )\wedge\widetilde{\omega}_{1}(\mu )
\ee
By means of (\ref{chon}), (\ref{ebol}) and (\ref{compota}) it follows that $d\Phi=d\ast \Phi=0$.

      The $G_2$ metrics (\ref{senio}) are fibered over the six dimensional metric
\begin{equation}
g_{6}=u\;d\mu ^{2}+\frac{(d\beta +H_{1})^{2}}{u}+g_{4}(\mu ),  \label{rel}
\end{equation}
which is K\"{a}hler with K\"{a}hler form
\begin{equation}
K=(d\beta +H_{1})\wedge d\mu +\widetilde{\omega}_{1}.  \label{chu3}
\end{equation}
The converse of these statements are also true. That is, for given a $G_{2}$ holonomy manifold $Y$ with a metric $g_{7}$ possessing a Killing vector that preserves the calibration forms $\Phi $ and $\ast \Phi $ and
such that the six-dimensional metric $g_{6}$ obtained from the orbits of the
Killing vector is K\"{a}hler, there exists a coordinate system in which
$g_{7}$ takes the form (\ref{senio}) being $g_{4}(\mu )$ a one-parameter
four-dimensional metric admitting a complex symplectic structure $\Omega $
and a complex structure $J_{1}$, being the quantities appearing in this
expression related by (\ref{areva}) and the conditions (\ref{chon}), (\ref{ebol}) and (\ref{compota}). This is the most involved part of the proofs and we refer the reader
to the original reference \cite{Apostolov}.

          The class metrics presented in this section include as particular cases the $G_2$ metrics which are dual to D6 branes wrapping an special lagrangian cycle and satisfying the strong supersymmetric conditions, i.e, the conditions for the supersymmetry generator to be a covariantly constant gauge spinor. These conditions will hold only if in  the string frame the IIA string metric has a Kahler internal part, which forces the $G_2$ dual metric to be in our class. Note that the Killing vector fields
preserve the metric and $\Phi$, thus it preserves $\ast \Phi$ and the whole $G_2$ structure. Another interesting fact is that
$$
\ast \Phi|_{M}=V(g_{4}),
$$
therefore the Kahler base $g_{4}$ is a coassociative submanifold . In the same way for fixed value of the coordinates of $g_4$ one obtains from (\ref{senio}) the three
dimensional metric
\begin{equation}
g_{3}=\frac{d\alpha^{2}}{\mu ^{2}}+ u\;d\mu ^{2}+
\mu \frac{d\beta^{2}}{u}.
\label{seniorita}
\end{equation}
defined on certain space $M_3$, and it follows that
$$
\ast \Phi|_{M_3}=V(g_3),
$$
therefore $M_3$ is an associative submanifold. These are calibrated submanifolds \cite{Lawson} and are supersymmetric from the physical point of view \cite{Strominger}.

\subsection{Uplifting to Spin(7) metrics}

    In the present subsection it will be shown that any of the $G_2$ holonomy metrics (\ref{senio}) described above can be lifted to an Spin(7) holonomy one by means of the lifting formula (\ref{mon}).  The two form $F=dA$ appearing in this formula must be closed. From the fact that 
$\Omega =\omega_{2}+i\omega_{3}$ is sympletic and that $dH_{2}=-\omega_{2}$  (see (\ref{chon})) the most natural anzatz is 
\be\lb{natans}
F=dA=-\omega_{3}.
\ee
The system (\ref{mon}) reduce in this case to
\be\lb{mondon}
\omega_3\wedge \Phi= d(e^{-4f})\wedge \ast \Phi.
\ee
From (\ref{dol}) it follows that the right hand of (\ref{mondon}) is
$$
d(e^{4f})\wedge\ast\Phi=d(e^{-4f})\wedge\bigg(\mu^2\widetilde{\omega}_{1}(\mu )\wedge d\mu\wedge(d\beta+H_1)+u \omega_{2}\wedge (d\alpha+H_2)\wedge d\mu
$$
\be\lb{levo}
+\omega_{3}\wedge (d\beta+H_1)\wedge (d\alpha+H_2)+\mu^2\widetilde{\omega}_{1}(\mu )\wedge\widetilde{\omega}_{1}(\mu )\bigg)
\ee
The left hand side is obtained from (\ref{dale}) and is
$$
\omega_3\wedge \Phi=\widetilde{\omega}_{1}(\mu )\wedge (d\alpha +H_{2})\wedge \omega_{3}+d\mu \wedge (d\beta
+H_{1})\wedge (d\alpha +H_{2}) \wedge \omega_{3}
$$
\be\lb{jesus}
+\mu \;\left( \;\omega_{2}\wedge (d\beta +H_{1})+u\omega%
_{3}\wedge d\mu \;\right)\wedge \omega_{3}
\ee
But from (\ref{compota}) we see that $\omega_{2}\wedge\omega_{3}=\widetilde{\omega}_{1}\wedge \omega_{3}=0$ and also that
$$u \omega_{3}\wedge \omega_{3}=\mu \widetilde{\omega}_{1}\wedge \widetilde{\omega}_{1}.$$ Taking into account these relations the formula (\ref{jesus}) is simplified to
\be\lb{smk}
\omega_3\wedge \Phi=d\mu \wedge\bigg( (d\beta
+H_{1})\wedge (d\alpha +H_{2} )\wedge \omega_{3}+\mu^2 \widetilde{\omega}_{1}\wedge \widetilde{\omega}_{1}\bigg) 
\ee
Equating (\ref{smk}) to (\ref{levo}) gives the equation
$$
d(e^{-4f})\wedge d\mu\wedge \left(\mu^2\widetilde{\omega}_{1}(\mu )\wedge(d\beta+H_1)-u \omega_{2}\wedge (d\alpha+H_2)\right)
$$
\be\lb{inna}
+d(e^{-4f})\wedge \left(\omega_{3}\wedge (d\beta+H_1)\wedge (d\alpha+H_2)+\mu^2\widetilde{\omega}_{1}(\mu )\wedge\widetilde{\omega}_{1}(\mu )\right)
\ee
$$
=d\mu \wedge\left(\omega_{3}\wedge (d\beta+H_1)\wedge (d\alpha+H_2)+\mu^2\widetilde{\omega}_{1}(\mu )\wedge\widetilde{\omega}_{1}(\mu )\right)
$$
The solution of this system is immediate. If $d(e^{-4f})=d\mu$ the two first terms of the left hand side vanishes and the two last ones equal to the right hand side. We choose then
$e^{-4f}=\mu+b$ and our Spin(7) metrics become 
\be\lb{uplift}
g_8=\frac{(dz+H_3)^2}{(\mu+b)^{3/2}}+(\mu+b)^{1/2}g_{7}
\ee
being $dH_3=-\omega_3$ and $g_7$ the $G_2$ holonomy metrics  described in the previous section.  Thus, we have found the Spin(7) metrics we were looking for. The reason for which the family is infinite is because the $G_2$ family over which are fibered is also infinite \cite{Apostolov}, \cite{Gaston}.

\section{Explicit Spin(7) examples}

   The local expression (\ref{uplift}) describe infinite family of Spin(7) metrics admitting $G_2$ reductions. In this section we found the metrics corresponding to known $G_2$ holonomy cases \cite{Gaston}. The lift in this case present no difficulties, but serves as warmup for the next section, in which less trivial examples will be worked out.

\subsection{Two different general solutions}

    Any of the Spin(7) metrics (\ref{uplift}) are fibrations over a $G_2$ holonomy metric $g_7$ of the type described in section 2.2, which are constructed in terms of solutions of the equations (\ref{areva})-(\ref{chon}). The 1-form $H_3$ satisfies $dH_3=-\omega_3$.  A simple example is obtained by assuming that the function $u$ defined in (\ref{compota}) does not vary when we move
on $M$ but depends on the coordinate $\mu$. Then the equation (\ref{ebol}) gives that $\widetilde{\omega}_1=(c\mu+d) \;\omega_1$ being $\omega_1$ independent on $\mu$. In addition $\widetilde{\omega}_1$ is closed on $M_4$ from where it follows that $d_4\omega_1=0$. This means that if one starts with an hyperkahler triplet $\omega_i$ of some hyperkahler manifold $M$ all
the conditions (\ref{senio})-(\ref{chon}) are solved except (\ref{compota}), which becomes then an algebraic equation defining $u$. The solution is $u=\mu \; (c\;\mu+d)^2$. Also $g_4(\mu)=(c\;\mu+d)\;\overline{g}_4$ being $\overline{g}_4$ the hyperkahler metric corresponding to $\omega_i$. The resulting 7-metrics (\ref{senio}) have the following expression
\begin{equation}
g_{7}=\frac{(d\alpha +H_{2})^{2}}{\mu^{2}}+%
\frac{(d\beta +H_{1})^{2}}{(c\mu^{2}+d)^2}+\mu^2 (c\mu+d)^2\;d\mu ^{2}+\mu (c\mu+d)\overline{g}_4.  \label{senio2}
\end{equation}%
Moreover the equations (\ref{chon}) are in this case
\begin{equation}
dH_{1}=\omega_{1},\qquad dH_{2}=-\omega_{2}. \label{chon2}
\end{equation}%
These metrics are usually well behaved away from the point $\mu=0$ or $\mu=-b$ if $b<0$.

   A second type of metrics are obtained with a function $u$ which depends on $\mu$ and also varies on $M$. This case is more difficult to deal with but still we will find
below several explicit examples. Consider as before an hyperkahler structure $\omega_i$ with its Ricci flat metric $\overline{g}_4$ and deform one of the Kahler two forms, say $\omega_1$,
to a new one $\widetilde{\omega}_1(\mu)$ of the form
\be\lb{anzatz}
\widetilde{\omega}_1(\mu)=\omega_1-d_4 d_4^c G,
\ee
being $G$ a function on $M\times R_{\mu}$. Then the compatibility conditions (\ref{areva0}) and (\ref{areva}) and (\ref{areva4}) are satisfied by (\ref{anzatz}).  Inserting (\ref{anzatz}) into the evolution equation (\ref{ebol}) gives
\be\lb{chus}
\partial_{\mu}^2 G=2\;u,
\ee
therefore $u$ is completely determined in terms of $G$. The equation for $G$ is found from (\ref{chon}) as follows. The relation
\be\lb{monsh}
\widetilde{\omega}_1(\mu)\wedge \widetilde{\omega}_1(\mu)=(\omega_1-d_4 d_4^c G)\wedge (\omega_1-d_4 d_4^c G)=\textit{M}(G)\omega_1\wedge \omega_1
\ee
defines a non linear operator $\textit{M}(G)$. This operator always exist as all the terms in (\ref{monsh})
are proportional to the volume form of $g_4(\mu)$. Then the insertion
of (\ref{chus}) into (\ref{chon}) gives
\be\lb{otra}
2\mu\; \textit{M}(G)=\partial_{\mu}^2 G,
\ee
which is the equation we were looking for. Also, from (\ref{ebol}) it follows that
\be\lb{di}
H_1=-d_4^c \partial_{\mu} G.
\ee
Note that in general, the metric tensor $g_{4}(\mu)$ in (\ref{senio}) is \emph{not} the hyperkahler metric $\overline{g}_4$ in general. If $K$ denote the Kahler potential corresponding
to  $\omega_1$ then the metric $g_{4}(\mu)$ is the one which corresponds to the modified Kahler potential $\overline{K}=K-G$. This metric is obviously Kahler, but
not necessarily hyperkahler. Equations (\ref{chus})-(\ref{di}) define a new family of $G_2$ metrics and all the objects defining the metric
are related essentially to a single function $G$ satisfying (\ref{otra}).

    To find the general solution of the previous equations is  extremely complicated because $\textit{M}(G)$ is a non linear operator.  The source of non linearity is
given by the term $d_4 d_4^c G\wedge d_4 d_4^c G$ in (\ref{monsh}). Nevertheless, there exist  special cases in which
\be\lb{bedford}
d_4 d_4^c G\wedge d_4 d_4^c G=0.
\ee
In these situations the operator $M(G)$ reduce to the linear operator \cite{Apostolov}
$$
\textit{M}(G)=1+\Delta_4 G
$$
being $\Delta_4$ the laplacian over the starting hyperkahler metric $\overline{g}_4$. In fact, the full system describing the $G_2$ geometry is linear in these cases. The condition (\ref{bedford}) is satisfied when the function $G$ is defined over a complex submanifold on the hyperkahler manifold $M$ \cite{Bedford}. This statement
means the following. The starting hyperkahler structure $\omega_i$ is obviously Kahler, thus $M$ is complex and parameterized in terms of certain complex coordinates
$(z_1, z_2, \overline{z}_1, \overline{z}_2)$ which diagonalize $J_1$. The equation (\ref{bedford}) will be satisfied when the function $G$ is of
the form $G=G(w, \overline{w})$ being $w$ a single complex function of the $z_i$ and $\overline{w}$ its complex conjugate. In particular, the equation (\ref{otra}) defining the $G_2$ geometry will be reduced to
\be\lb{otra2}
2\mu(1+\Delta_4 G)=\partial_{\mu}^2 G.
\ee
This is an important simplification, although the task of finding solutions of a Laplace equation in a curved space is not easy in general. In the next section we will find an explicit solution by taking the Eguchi-Hanson gravitational instanton as our initial hyperkahler structure.

\subsection{Simple known examples}

    The solution generating techniques described in the previous subsection require an initial hyperkahler structure. 
The simplest hyperkahler manifold is $R^4$ with its flat metric
$g_4=dx^2+dy^2+dz^2+d\varsigma^2$ and with the closed hyperkahler
triplet
$$
\omega_1=d\varsigma\wedge dy- dz\wedge dx,\qquad
\omega_2=d\varsigma\wedge dx- dy\wedge dz,\qquad
\omega_3=d\varsigma\wedge dz- dx\wedge dy.
$$
This innocent looking case is indeed very interesting. Let us construct the metrics (\ref{senio2}) corresponding to this structure. The forms $H_i$ such 
that $dH_i=\omega_i$ are simply
\be\lb{susu}
H_1=-x dz + y d\varsigma,\qquad 
H_2=-y dz+x d\varsigma,\qquad H_3=-y dx+ z d\varsigma
\ee
and by selecting $c=1$ and $d=0$ in (\ref{senio2}) the resulting $G_2$ metric is \be\lb{plani2} g_7=
\frac{(d\alpha-x dz + y d\varsigma)^2}{\mu^2}+ \frac{(d\beta-y
dz-x d\varsigma)^2}{\mu^2} +\mu^4\;d\mu^2
+\mu^2\;(\;dx^2+dy^2+dz^2+d\varsigma^2\;). \ee 
The metrics (\ref{plani2}) have been already obtained in the physical literature
\cite{Stelle} and are interpreted in terms of domain wall configurations . Even in this simple case and though the base 4-metric has trivial holonomy, it has been shown that (\ref{plani2}) 
is irreducible and has holonomy exactly $G_2$, not a subgroup.

  Turning on the attention to the second ramification, a possible choice of complex coordinates
for $R^4$ is $z_1=x+i y$, $z_2=z+i\varsigma$ and complex conjugates. If the functional dependence the function $G$ is assumed to be $G=G(\mu, z_1, \overline{z}_1)$ then
the operator $\textit{M}(G)$ reduce to the
laplacian operator in flat space \be\lb{air} G''+\mu(\partial_{xx}G +
\partial_{yy}G)=2\mu. \ee
The separable solutions in
the variable $\mu$ are of the form
$$
G=\frac{1}{3}\mu^3+V(x,y)K(\mu).
$$
By introducing $G=G(\mu, x,y)$ into (\ref{air}) it follows that
$K(\mu)$ and $V(x,y)$ are solutions of the equations \be\lb{m}
K''(\mu)=p\;\mu\; K(\mu),\qquad \partial_{xx}V +
\partial_{yy}V + p\;V=0, \ee being $p$ a parameter. By defining the
$\widetilde{\mu}=\mu/p^{1/3}$ the first of the equations
(\ref{m}) reduce to the Airy equation. The second is reduced to find
eigenfunctions of the two dimensional Laplace operator, which is a
well known problem in electrostatics. For $p>0$ periodical solutions
are obtained and for $p<0$ there will appear exponential solutions.

    A simple example is obtained with the eigenfunction $V=q\;\sin(p\;x)$, being
$q$ a constant. A solution of the
Airy equation is given by
$$
K=Ai(\widetilde{\mu})=\frac{1}{3}\widetilde{\mu}^{1/2}(J_{1/3}(\tau)
+ J_{-1/3}(\tau)),\qquad \tau=i \frac{2\;\mu^{3/2}}{3\;p^{1/2}}.
$$
Then the function $G$ is
$$
G=\frac{1}{3}\mu^3+ q\;\sin(p\;x) Ai(\frac{\mu}{p^{1/3}}),
$$
From (\ref{di}) it is obtained that
\be\lb{kdjd}
H_1=-p\;q Ai(\widetilde{\mu})' \cos(p\;x) dy,\qquad u=\mu (1+ p\;q\; Ai(\widetilde{\mu})\;\sin(p\;x)).
\ee
$$
g_{4}(\mu)=\frac{u}{\mu} (dx^2+ dy^2)+ dz^2+ d\varsigma^2
$$
In terms of the quantities defined above the generic $G_2$ holonomy metric (\ref{senio}) becomes \cite{Apostolov} \be\lb{airy}
g_{7}=\frac{(d\chi-x dz+y d\varsigma)^2}{\mu^2}
+\frac{(d\upsilon-p\;q Ai(\widetilde{\mu})' \cos(p\;x) dy)^2}{H}
+\mu\;(\;H dx^2+H dy^2+ dz^2+ d\varsigma^2\;) +\mu^2 H \;
d\mu^2, \ee
where the function $H(\mu,x,y)=(1+ p\;q\;
Ai(\widetilde{\mu})\;\sin(p\;x))$ has been introduced. As before, the holonomy is \emph{exactly} $G_2$ \cite{Apostolov}.

       It is straightforward to construct from (\ref{airy}) or (\ref{plani2}) a pair of holonomy $Spin(7)$ metrics, which 
are obtained from (\ref{uplift}) and (\ref{susu}). The result is
\be\lb{uplift2}
g_8=\frac{(dz+ydx-zd\varsigma)^2}{(\mu+b)^{3/2}}+(\mu+b)^{1/2}g_{7}
\ee
being $g_7$ any of (\ref{airy}) or (\ref{plani2}). The curvature tensor is irreducible for these metrics
and the holonomy is not reduced to a subgroup.

\section{Two fibrations over the Eguchi-Hanson gravitational instanton}

\subsection{The Eguchi-Hanson metric as an ALE space}

    In this section the solutions of the two ramifications described above will be worked in the situation in which the Eguchi-Hanson gravitational instanton
is the initial hyperkahler metric \cite{Eguchi}. As is well known, this metric is preserved by an isometry which also preserve its Kahler forms $\omega_i$, namely 
\[
{\cal L}_{K}\omega_{1}={\cal L}_{K}\omega_{2}={\cal L}_{K}%
\omega_{3}=0
\]%
being $K$ the corresponding Killing vector. Such Killing vector $K$ is called tri-holomorphic and it also preserve the complex structures $J_i$
defined by the Kahler forms and the metric, thus it is also tri-hamiltonian. For any 4-dimensional hyperkahler structure with this property there exist
a local system of coordinates in which  $K=\partial _{t}$ and for which the metric takes generically the
Gibbons-Hawking form \cite{Gibbhawk}
\begin{equation}
g=V^{-1}(dt+A)^{2}+Vdx_{i}dx_{j}\delta ^{ij},  \label{ashgib}
\end{equation}%
with a 1-form $A$ and a function $V$ satisfying the linear system of
equations
\begin{equation}
\nabla V=\nabla \times A.  \label{Gibb-Hawk}
\end{equation}%
In addition the hyperkahler triplet in this coordinates is given by
\[
\omega_{1}=(dt+A)\wedge dx-Vdy\wedge dz
\]%
\begin{equation}
\omega_{2}=(dt+A)\wedge dy-Vdz\wedge dx  \label{transurop3}
\end{equation}%
\[
\omega_{3}=(dt+A)\wedge dz-Vdx\wedge dy
\]%
which is actually $t$-independent. The Eguchi-Hanson solution corresponds to two monopoles on the $z$ axis. Without
losing generality, it can be considered that the monopoles are located in
the positions $(0,0,\pm c)$. The potentials for this configurations are
\[
V=\frac{1}{r_{+}}+\frac{1}{r_{-}},\qquad A=A_{+}+A_{-}=\left( \frac{z_{+}}{%
r_{+}}+\frac{z_{-}}{r_{-}}\right) d\arctan (y/x),\qquad r_{\pm
}^{2}=x^{2}+y^{2}+(z\pm c)^{2}.
\]%
The resulting metric (\ref{ashgib}) in
Cartesian coordinates is
\begin{equation}
g=\left( \frac{1}{r_{+}}+\frac{1}{r_{-}}\right) ^{-1}\left( \;d\tau +\left(
\frac{z_{+}}{r_{+}}+\frac{z_{-}}{r_{-}}\right) \;d\arctan (y/x)\;\right)
^{2}+\left( \;\frac{1}{r_{+}}+\frac{1}{r_{-}}\;\right)
(\;dx^{2}+dy^{2}+dz^{2}\;),  \label{eguchi}
\end{equation}%
where $z_{\pm }=z\pm c$. In order to recognize the Eguchi-Hanson metric in
its standard form it is convenient to introduce a new parameter $a^{2}=8c,$
and the elliptic coordinates defined by \cite{Prasad}
\[
x=\frac{r^{2}}{8}\sqrt{1-(a/r)^{4}}\sin \varphi \cos \theta ,\quad y=\frac{%
r^{2}}{8}\sqrt{1-(a/r)^{4}}\sin \varphi \sin \theta ,\quad z=\frac{r^{2}}{8}%
\cos \varphi .
\]%
It is not difficult to check that in these coordinates
\[
r_{\pm }=\frac{r^{2}}{8}\left( 1\pm \left( a/r\right) ^{2}\cos \varphi
\right) ,\qquad z_{\pm }=\frac{r^{2}}{8}\left( \cos \varphi \pm
(a/r)^{2}\right) ,\qquad V=\frac{16}{r^{2}}\left( 1-(a/r)^{4}\cos
^{2}\varphi \right) ^{-1},
\]%
\[
A=2\;\left( 1-(a/r)^{4}\cos ^{2}\varphi \right) ^{-1}\;\left(
1-(a/r)^{4}\right) \;\cos \varphi \;d\theta ,
\]%
and, with the help of these expressions, it is found
\begin{equation}
g=\frac{r^{2}}{4}\left( \;1-(a/r)^{4}\;\right) \;(\;d\theta +\cos \varphi
d\tau \;)^{2}+\left( \;1-(a/r)^{4}\;\right) ^{-1}\;dr^{2}+\frac{r^{2}}{4}%
\;(\;d\varphi ^{2}+\sin ^{2}\varphi d\tau \;)  \label{ego}
\end{equation}%
This is actually a more familiar expression for the Eguchi-Hanson instanton,
indeed. Its isometry group is $U(2)=U(1)\times SU(2)/{\bf Z}_{2}$. The holomorphic Killing vector is $%
\partial _{\tau }$. This space is asymptotically locally Euclidean (ALE),
which means that it asymptotically approaches the Euclidean metric; and
therefore the boundary at infinity is locally $S^{3}$. However, the
situation is rather different in what regards its global properties. This
can be seen by defining the new coordinate
\[
u^{2}=r^{2}\left( 1-(a/r)^{4}\right) 
\]%
for which the metric is rewritten as%
\begin{equation}
g=\frac{u^{2}}{4}\;(\;d\theta +\cos \varphi d\tau \;)^{2}+\left(
\;1+(a/r)^{4}\;\right) ^{-2}\;du^{2}+\frac{r^{2}}{4}\;(\;d\varphi ^{2}+\sin
^{2}\varphi d\tau \;).  \label{ego2}
\end{equation}%
The apparent singularity at $r=a$ has been moved now to $u=0$. Near the
singularity, the metric looks like 
\[
g\simeq \frac{u^{2}}{4}\;(\;d\theta +\cos \varphi d\tau \;)^{2}+\frac{1}{4}%
du^{2}+\frac{a^{2}}{4}\;(\;d\varphi ^{2}+\sin ^{2}\varphi d\tau \;), 
\]%
and, at fixed $\tau $ and $\varphi ,$ it becomes 
\[
g\simeq \frac{u^{2}}{4}\;d\theta ^{2}+\frac{1}{4}du^{2}. 
\]%
This expression \textquotedblleft locally\textquotedblright\ looks like the
removable singularity of ${\bf R}^{2}$ that appears in polar coordinates.
However, for actual polar coordinates, the range of $\theta $ covers from $0$
to $2\pi $, while in spherical coordinates in ${\bf R}^{3},$ $0\leq \theta
<\pi $. This means that the opposite points on the geometry turn out to be
identified and thus the boundary at infinite is the lens space $S^{3}/{\bf Z}%
_{2}$. 

\subsection{The first type of metrics}

  The task to find the $G_2$ metrics (\ref{senio2})  that correspond to the Eguchi-Hanson instanton was already solved in \cite{Gaston}. 
 The explicit expression of the 1-forms $H_i$ satisfying  $dH_i=\omega_i$, being $\omega_i$ the Kahler forms (\ref{transurop3}), is the following
\begin{equation}
H_{1}=-x\;d\tau +\left( \log (r_{+}+z_{+})+\log (r_{-}+z_{-})\right)
dy-2a\;x\;d\arctan (y/x),  \label{ssssrrrr}
\end{equation}%
\begin{equation}
H_{2}=+y\;d\tau \;+\left( \log (r_{+}+z_{+})+\log (r_{-}+z_{-})\right)
\;dx+2a\;y\;d\arctan (y/x),  \label{integrals}
\end{equation}%
\begin{equation}
H_{3}=-zd\tau -a\;(\;r_{+}+r_{-}\;)\;d\arctan (y/x).  \label{rrrrssss}
\end{equation}
These expressions are defined up to a redefinition by a total differential and, together with (\ref{senio2}), define the following $G_2$ metric
\begin{equation}
g_{7}=\frac{(d\alpha +H_{2})^{2}}{\mu^{2}}+%
\frac{(d\beta +H_{1})^{2}}{(c\mu^{2}+d)^2}+\mu^2 (c\mu+d)^2\;d\mu ^{2}+\mu (c\mu+d)\overline{g}_4\label{senio3}
\end{equation}%
where $\overline{g}_4$ the Eguchi-Hanson metric. In addition, a Spin(7) holonomy metric is obtained directly from (\ref{senio3}) and (\ref{uplift}). 

\subsection{The new examples}

     In this subsection we construct a family of $G_2$ and Spin(7) metrics fibered over the Eguchi-Hanson instanton which, to our knowledge, have not been
previously considered.  For this it is convenient to define a new radial coordinate $\rho=r^2/4$ for Eguchi-Hanson metric (\ref{ego}) while keeping the angular coordinates unchanged. In terms of 
these coordinates (\ref{ego}) takes the form
\be\lb{Ehan}
g_{EH}=\frac{\rho}{\rho^2-a^2}d\rho^2+\rho(\sigma_1^2+\sigma_2^2)+\frac{\rho^2-a^2}{\rho}\sigma_3^2
\ee
being 
$$
\sigma_1=\frac{1}{2}(\cos\theta d\varphi+\sin\theta \sin\varphi d\tau),
$$
$$
\sigma_2=\frac{1}{2}(-\sin\theta d\varphi+\cos\theta \sin\varphi d\tau),
$$
$$
\sigma_3=\frac{1}{2}(d\theta+\cos\varphi d\tau).
$$
The Kahler forms are given by
\be\lb{ijo}
\omega_i=e^0 \wedge e^i-\epsilon_{ijk}\;e^j\wedge e^k,
\ee
being $e^i$ the tetrad basis
$$
e^0=\sqrt{\frac{\rho}{\rho^2-a^2}}d\rho,\;\;\; e^{1,2}=\sqrt{\rho}\sigma_{1,2},\;\;\; e^{3}=\sqrt{\frac{\rho^2-a^2}{\rho}}\sigma_{3}.
$$
As usual, the hyperkahler structure (\ref{Ehan}) and (\ref{ijo}) will be starting point for constructing a $G_2$ holonomy metric. This is achieved with the help of  a function $G$ satisfying the laplace type equation (\ref{otra2}) together with the condition (\ref{bedford}). As the last condition implies that $G$ is defined on a complex submanifold of the hyperkahler space, it is necessary to find a complex coordinate system for (\ref{Ehan}). A well known coordinate system is 
the one which diagonalize the complex structure $J_3$ corresponding to the Kahler form $\omega_3$. These coordinates are \cite{Gibbonsviejo}
$$
z_1=(\rho^2-a^2)^{1/4}\cos(\frac{\varphi}{2}) \exp(i \frac{\theta+\tau}{2}),
$$
\be\lb{complexcord}
z_2=(\rho^2-a^2)^{1/4}\sin(\frac{\varphi}{2}) \exp(i \frac{\theta-\tau}{2}).
\ee
The hyperkahler metric (\ref{Ehan}) is expressed in this coordinates as
\be\lb{comcor}
g_{1\overline{1}}=\frac{\rho^2|z_2|^2+ \eta^2 |z_1|^2}{\rho \eta^2},
\ee
\be\lb{comcor2}
g_{2\overline{2}}=\frac{\rho^2|z_1|^2+ \eta^2 |z_2|^2}{\rho \eta^2},
\ee
\be\lb{comcor3}
g_{1\overline{2}}=\frac{\eta^2-\rho^2}{\rho \eta^2}z_2 \overline{z}_1,
\ee
which is symmetric under the interchange $z_1\leftrightarrow z_2$. We have denoted $\eta=|z_1|^2+ |z_2|^2=\sqrt{\rho^2-a^2}$.
The advantage of consider this coordinates are clear when calculating the laplacian
$$
\Delta_{EH}=\frac{1}{\sqrt{\det(g)}}\partial_i(\sqrt{\det(g)}g^{ij}\partial_j).
$$
In this coordinates $\det(g)=1$ and the inverse metric is simply
$$
g^{1\overline{1}}=g_{2\overline{2}},\qquad g^{2\overline{2}}=g_{1\overline{1}},\qquad g^{1\overline{2}}=-g_{2\overline{1}}.
$$
Moreover, after certain calculation is obtained that
$$
\partial_1(g^{1\overline{1}})=-\partial_2(g^{2\overline{1}}),\qquad\partial_{\overline{1}}(g^{\overline{1}1})=-\partial_{\overline{2}}(g^{\overline{2}1}).
$$
The last equalities are more easily checked with \textit{Mathematica} than by hand. From them 
it follows that the action of the laplacian acting on a function $U(z_1, \overline{z}_1)$ is simply
\be\lb{simpleform}
\Delta_{EH}U=g^{1\overline{1}}\partial_1 \partial_{\overline{1}}U.
\ee
As we have explained, equation (\ref{otra}) will become linear if the dependence of $G$ with respect to the complex coordinates 
is $G=G(w, \overline{w})$ being $w=w(z_1, z_2)$ an holomorphic function of $z_1$ and $z_2$ and $\overline{w}$ its complex conjugate. If additionally  $w(z_1, z_2)=z_1$ 
then the action of the laplacian will be simply (\ref{simpleform}).  For this reason we assume that $G=G(\mu, z_1, \overline{z}_1)$. The equation (\ref{otra2}) is simplified
with this anzatz to 
\be\lb{otra3}
\mu (1+g^{1\overline{1}}\partial_1 \partial_{\overline{1}}G)=\partial^2_{\mu}G.
\ee
But the component $g^{1\overline{1}}$ is a function of $z_2$ and $G$, by  our assumption, is not. This observation together with (\ref{otra3}) imply that
\be\lb{cuto}
\partial_1 \partial_{\overline{1}}G=0,\qquad  \partial^2_{\mu}G=\mu.
\ee
The most general solution of (\ref{cuto}) is simply
\be\lb{moge}
G=\frac{\mu^3}{3}+\mu\left(F(z_1)+\overline{F}(\overline{z}_1)\right)+ H(z_1)+\overline{H}(\overline{z}_1),
\ee
being $F$ and $H$ functions on the complex coordinate $z_1$ and $\overline{F}$ and $\overline{H}$ their complex conjugated.

   The function $G$ found above determines completely a family of special holonomy metrics given by (\ref{senio}) and (\ref{uplift}). The fiber quantities in these expressions are obtained
 as follows. From (\ref{chus}) and (\ref{moge}) it follows that
$2\;u=\mu$. The expression of the exterior derivatives over the Eguchi-Hanson manifold in our complex coordinates is 
$$
d_4=\partial_{z_i}dz^i+\partial_{\overline{z}_i}d\overline{z}^i,\qquad
d_4^c=i\partial_{z_i}dz^i-i\partial_{\overline{z}_i}d\overline{z}_i
$$
and their action over (\ref{moge}) gives
$$
d_4 d_4^c G=0.
$$
From (\ref{anzatz}) and the last equation it is obtained that $\widetilde{\omega}_1(\mu)=\omega_1$
which means that $g_4=\overline{g}_4$ is the Eguchi-Hanson metric $g_{EH}$. Note that this equality is accidental, for other initial hyperkahler structure
it may not hold. From (\ref{di}) it is obtained that
\be\lb{di2}
H_1=i\mu (F^{'} dz^1-\overline{F}^{'}d\overline{z}_1)+i(H^{'} dz^1-\overline{H}^{'}d\overline{z}_1)=\Im \left((\mu F^{'}+ H^{'})dz_1\right)
\ee
and is clear that it takes real values. Here ' means the derivative with respect to the argument. 
The corresponding $G_2$ and Spin(7) metrics are easily constructed from (\ref{senio}) and (\ref{uplift}) and the quantities defined above, the result is
\begin{equation}
g_{7}=\frac{\mu^2}{2}d\mu^{2}+%
(d\beta +\Im (F) d\mu)^{2}+\frac{(d\alpha +H_{2})^{2}}{\mu^2}+ \mu \; g_{EH}.  \label{seniol}
\end{equation}%
for the $G_2$ holonomy metrics and
\be\lb{upliftl}
g_8=\frac{(dz+H_3)^2}{(\mu+b)^{3/2}}+(\mu+b)^{1/2}\bigg(\frac{\mu^{2}}{2} d\mu ^{2}+
2(d\beta +\Im (F) d\mu)^{2}+\frac{(d\alpha +H_{2})^{2}}{\mu^{2}}+ \mu \; g_{EH}\bigg),
\ee
for the Spin(7) holonomy ones. Here $d\beta$ has been redefined by adding a total differential and $H_1$ and $H_2$ are the forms (\ref{ssssrrrr}) or (\ref{integrals}). 

  A simple inspection shows that neither of the special holonomy metrics constructed above has a signature change problem.
By defining the proper coordinate $\tau=\mu^2/2$ the $G_2$ metric (\ref{seniol}) becomes \begin{equation}
g_{7}=d\tau^{2}+%
\frac{(\tau d\beta +\Im (F) d\tau)^{2}}{\tau}+\frac{(d\alpha +H_{2})^{2}}{\tau}+ \tau^{1/2} g_{EH},  \label{seniolo}
\end{equation}%
and we see from the square root that $\tau$ take positive values and there is no change in the signature. Also, by selecting $b=0$ in the Spin(7) metric
and defining $\eta=\mu^{9/4}$ it is obtained the following expression 
\be\lb{upliftlo}
g_8=d\eta^2+
\frac{(\eta^{5/9} d\beta +\Im (F) d\eta)^{2}}{\eta^{8/9}}+\frac{(dz+H_3)^2}{\eta^{2/3}}+\frac{(d\alpha +H_{2})^{2}}{\eta^{2/3}}+ \eta^{2/3} g_{EH}
\ee
for (\ref{upliftl}) and in this case the powers of $\eta$ are all even and there is not signature change as $\eta$ goes from positive to negative values.

   The class of metrics (\ref{upliftl}) and (\ref{seniol}) depend on an arbitrary choice of an holomorphic function $F(z_1)$. This is the only freedom
to construct them. In fact both metrics arise as an $R_{\beta}$-fibration and  their quotient by the $\beta$-isometry gives the same 
6 or 7 dimensional metric. The function $F$ indicates the way the lift of these 6 or 7 dimensional metrics
to a $G_2$ or Spin(7) holomy one is performed. Therefore (\ref{upliftl}) and (\ref{seniol}) describe an infinite family of special holonomy metrics.

\section{Discussion}

    The main result of the present work is the proof that \emph{any} metric of $G_2$ holonomy  with an isometry action preserving the metric and the calibration 3 and 4-forms in such a way that the quotient of the $G_2$ structure by this action is \emph{Kahler} admits a lift to a closed Spin(7) structure. As the initial 7-metric always possess two commuting Killing vectors, the underlying Spin(7) holonomy metric will possess three commuting Killing vectors. All these special holonomy manifolds are non compact in view of the description given in \cite{Apostolov} for the $G_2$ case. Additionally, we have constructed several families of special holonomy metrics and we remark (\ref{upliftlo}) and (\ref{seniol}) which, in our opinion, are new examples. We were unable to show wether or not a $G_2$ holonomy metric obtained by the quotient of a closed Spin(7) structure by an isometry preserving it admits a further reduction to a Kahler 6-dimensional structure. This is an open question.
    
     Although our results are formulated in mathematical terms, they have physical consequences. The class of $G_2$ holonomy metrics admitting Kahler reductions include the examples which realize a dual description of configurations of D6 branes wrapping a special lagrangian cycle in a non compact CY 3-fold and satisfying the strong supersymmetry conditions, which are the conditions that convert the supersymmetry generator into a covariantly constant gauge spinor.  The presence of the D6 branes is due to sources for the RR two form F, that is, dF $\sim\;N \;\delta$. The dual description is achieved in terms of a 11-dimensional background is the direct sum of the $G_2$ metric plus the flat Minkowski metric in four dimensions. As our results imply that any of such $G_2$ metrics can be lifted to one of Spin(7) holonomy, one can construct as well a purely geometrical background in eleven dimensions which is the direct sum of the Spin(7) metric plus the flat Minkowski metric in three dimensions. The usual Kaluza-Klein reduction along one of the isometries will give a new IIA background with a non trivial dilaton and a self-dual RR two form F$^{'}$ in such a way that the 10-dimensional metric in the string frame contains a internal part conformal to the $G_2$ metric. It seems reasonable to postulate that delta sources for the RR two form F$^{'}$ will be present as well,  and the new IIA background will describe a configuration of D6 branes wrapping a coassociative 4-cycle of the $G_2$ holonomy manifold. Unfortunately, this statement concerning the singularities of the RR two form does not follow directly from our analysis. One can argue that there may be no delta singularities for the new configuration but instead, the new RR two form is non trivial due to a bad behavior at infinite. We believe that this will not the case when the initial configuration corresponds to D6 branes, but we do not have a proof. This is an important problem for the following reason. The IIA backgrounds described above are completely determined in terms of the $G_2$ holonomy metric. If a given $G_2$ metric is determining completely a pair of such D6 brane configurations, then there may be a duality connecting the corresponding supersymmetric theories, and the link is provided by the $G_2$ structure. In our opinion, this possibility deserves further attention, as this duality will connect theories in three and four dimensions.
\\

{\bf Acknowledgement:} O. S is partially supported by CONICET, Argentina and by the Science Foundation Ireland Grant No. 06/RFP/MAT050.


\begin{thebibliography}{99}

\bibitem{Berger} M. Berger Bull.Soc.Math.Fr. 83, 279.
\bibitem{Bryant} R. Bryant Annals Math. 126, 525; R.Bryant and S. Salamon Duke Math.J.58 (1989) 829.
\bibitem{Gibbons}  G. Gibbons, D. Page and C. Pope Commun.Math.Phys.127 (1990) 529.
\bibitem{Joyce} D. Joyce J.Diff.Geom. 43, 2;  J.Diff.Geom. 43, 329.
\bibitem{Kowaleski} A. Kovalev J. Reine Angew. Math. 565 (2003), 125. 
\bibitem{novo1} M. Cvetic, G.W. Gibbons, H. Lu and C.N. Pope Nucl.Phys. B620 (2002) 29; J.Geom.Phys. 49 (2004) 350; Annals Phys. 310 (2004) 265; Phys.Rev. D65 (2002) 106004.
\bibitem{novo3} Y. Konishi and M. Naka Class.Quant.Grav. 18 (2001) 5521.
\bibitem{novo5} G. Etesi Phys.Lett. B521 (2001) 391.
\bibitem{Berglund} A.Brandhuber and P.Berglund, Nucl.Phys. B641 (2002) 351.
\bibitem{novo7} A. Misra J.Math.Phys. 47 (2006) 033504; A. Franzen, P. Kaura, A. Misra and R. Ray Fortsch.Phys. 54 (2006) 207.
\bibitem{Floratos} I. Bakas, E. Floratos, A. Kehagias, Phys.Lett. B445 (1998) 69.
\bibitem{Mahapatrah} K.Behrndt,G. Dall'Agata, D. Lust and S. Mahapatra, JHEP 0208 (2002) 027;
K.Behrndt, Nucl.Phys. B635 (2002) 158.
\bibitem{novo8} Z.W. Chong, M. Cvetic, G.W. Gibbons, H. Lu, C.N. Pope and P. Wagner Nucl.Phys. B638 (2002) 459.
\bibitem{novo88} L. Anguelova and C. Lazaroiu JHEP 0301 (2003) 066; JHEP 0210 (2002) 038.
\bibitem{novo89} O.P.Santillan Nucl.Phys. B660 (2003) 169.
\bibitem{Bilal} A.Bilal, J.Derendinger and K.Sfetsos,  Nucl.Phys. B628 (2002) 112.
\bibitem{novo9} M. Cvetic, G.W. Gibbons, H. Lu and C.N. Pope Class.Quant.Grav. 20 (2003) 4239.
\bibitem{novo10} A. Brandhuber, J. Gomis, S. Gubser and S. Gukov Nucl.Phys. B611 (2001) 179.
\bibitem{Stelle} G.Gibbons, H. Lu, C. Pope and K. Stelle Nucl.Phys. B {\bf 623} (2002) 3.
\bibitem{Apostolov} V. Apostolov and S. Salamon Comm. Math. Phys {\bf 246} (2004) 43.
\bibitem{Gaston} G. Giribet and O. Santillan Commun.Math.Phys. 275 (2007) 373.
\bibitem{Ego} O.P.Santillan Phys.Rev. D73 (2006) 126011.
\bibitem{Atiyah}  M. Atiyah, J. Maldacena and C. Vafa J.Math.Phys. 42 (2001) 3209.
\bibitem{Acharya} B. Acharya "On Realising N=1 Super Yang-Mills in M theory" hep-th/0011089.
\bibitem{Vafa} M.Aganagic and C.Vafa  JHEP 0305 (2003) 061; "$G_2$ Manifolds, Mirror Symmetry and Geometric Engineering" hep-th/0110171.
\bibitem{Minasian} P. Kaste, R. Minasian, M. Petrini and A. Tomasiello HEP 0209 (2002) 033.
\bibitem{Pap} S. Shatashvili and C. Vafa Selecta Math. 1 (1995) 347; G.Papadopoulos and K. Townsend Phys.Lett. B357 (1995) 300.
\bibitem{Eguchi}  T. Eguchi and A. Hanson, Phys. Lett. {\bf B74} (1978).
249; Annals Phys. {\bf 120} (1979) 82; Gen.Rel.Grav. {\bf 11} (1979) 315.
\bibitem{Gibbonsviejo} G. Gibbons and C. Pope Comm.Math.Phys {\bf 66} (1979) 267.
\bibitem{Prasad} M. Prasad Phys. Lett B 83 (1979) 310.
\bibitem{Chiossi} S. Chiossi and S. Salamon Proc. conf. Differential Geometry Valencia 2001.
\bibitem{Bedford} E. Bedford and M. Kalka Commun. Pure. Appl. Math {\bf 30} (1977) 543.
\bibitem{Kobayashi} R. Kobayashi Adv. Stud. Pure. Math 18 II (1990) 137.
\bibitem{Lawson} R. Harvey and H. B. Lawson Acta Math.148 (1982) 47.
\bibitem{Strominger} K.Becker, M. Becker and A. Strominger Nucl.Phys. B456 (1995) 130.
\bibitem{Gibbhawk} G. Gibbons and S. Hawking Phys. Lett B 78 (1978) 430.
\bibitem{Becker0} M. Atiyah and E. Witten JHEP 0606 (2006) 051.
\bibitem{Becker}  K. Becker JHEP 0105 (2001) 003.
\bibitem{Becker2} Melanie Becker, Dragos Constantin, S. James Gates, Jr., William D. Linch III, Willie Merrell and J. Phillips Nucl.Phys. B683 (2004) 67;
D. Constantin Nucl.Phys. B706 (2005) 221.
\bibitem{Becker5} K. Becker and M. Becker Nucl.Phys. B477 (1996) 155.
\bibitem{Becker6} B. Acharya and S. Gukov JHEP 0209 (2002) 047.
\bibitem{Becker61} A. Belhaj J.Phys. A36 (2003) 4191.
\bibitem{Becker62} A. Belhaj, M.P. Garcia del Moral, A. Restuccia, A. Segui and J.P. Veiro "The Supermembrane with Central Charges on a G2 Manifold" arXiv:0803.1827;
A. Belhaj, L. Boya and A. Segui "Relation Between Holonomy Groups in Superstrings, M and F-theories" arXiv:0806.4265.
\bibitem{Becker66} U. Gursoy, C. Nunez and M. Schvellinger JHEP 0206 (2002) 015.
\bibitem{Becker7} K.Becker, M.Becker, D.R.Morrison, H.Ooguri, Y.Oz and Z.Yin Nucl.Phys. B480 (1996) 225.
\bibitem{Becker10} G.W. Gibbons and G. Papadopoulos Commun.Math.Phys. 202 (1999) 593.
\bibitem{Faya} A. Fayazzuddin and T. Hussein Phys.Rev. D75 (2007) 065017.
\bibitem{Becker12} Sergei Gukov, Shing-Tung Yau, Eric Zaslow "Duality and Fibrations on $G_2$ Manifolds" hep-th/0203217.
\bibitem{Becker13} S. Frolov and A. Tseytlin Nucl.Phys. B632 (2002) 69; H. Lu, C.N. Pope, K.S. Stelle, P.K. Townsend JHEP 0410 (2004) 019; JHEP 0507 (2005) 075.
\bibitem{Becker14} S. Hartnoll and C. Nunez JHEP 0302 (2003) 049.
\bibitem{Becker15} J.D. Edelstein, A. Paredes and A.V. Ramallo JHEP 0301 (2003) 011.
\bibitem{Medeiros} J. de Boer, P. Medeiros, S. El-Showk and A. Sinkovics JHEP 0802 (2008) 012; Class.Quant.Grav.25 (2008) 075006.
\bibitem{Zabzine} G. Bonelli and M. Zabzine JHEP 0509 (2005) 015; G. Bonelli, A. Tanzini and M. Zabzine Adv.Theor.Math.Phys. 10 (2006) 239; 
JHEP 0703 (2007) 023.
\bibitem{Becker16} J. Gutowski and G. Papadopoulos Nucl.Phys. B615 (2001) 237.
\bibitem{Becker17} B. Acharya "Confining Strings from $G_2$ holonomy spacetimes" hep-th/0101206.
\bibitem{Becker18} P. Svrcek and E. Witten JHEP 0606 (2006) 051.
\end{thebibliography}
\end{document}